\documentclass[11pt]{article}
\usepackage{graphicx}

\usepackage{psfig}

\newcommand{\BABARPubYear}    {06}

\newcommand{\BABARProcNumber} {100}
\newcommand{\SLACPubNumber} {12163}

\usepackage{xspace}
\usepackage{relsize}

\def\babar{\mbox{\slshape B\kern-0.1em{\smaller A}\kern-0.1em
    B\kern-0.1em{\smaller A\kern-0.2em R}}}
 
\newcommand{\mev}{\ensuremath{\mathrm{\,Me\kern -0.1em V}}\xspace}
\newcommand{\mevcc}{\ensuremath{{\mathrm{\,Me\kern -0.1em V\!/}c^2}}\xspace}
\newcommand{\gevcc}{\ensuremath{{\mathrm{\,Ge\kern -0.1em V\!/}c^2}}\xspace}
\newcommand{\gevc}{\ensuremath{{\mathrm{\,Ge\kern -0.1em V\!/}c}}\xspace}
\newcommand{\gev}{\ensuremath{\mathrm{\,Ge\kern -0.1em V}}\xspace}
\def\invfb   {\ensuremath{\mbox{\,fb}^{-1}}\xspace}
\mathchardef\Upsilon="7107
\def\Y#1S{\ensuremath{\Upsilon{(#1S)}}\xspace}
\def\FourS {\Y4S}
\mathchardef\Xi="7104
\mathchardef\Lambda="7103
\mathchardef\Sigma="7106
\mathchardef\Omega="710A

\def\pip   {\ensuremath{\pi^+}\xspace}
\def\pim   {\ensuremath{\pi^-}\xspace}

\def\ra    {\ensuremath{\rightarrow}\xspace}
 
\def\lcpl  {\ensuremath{\Lambda^+_c}\xspace}
\def\brc   {\ensuremath{{\cal B}}\xspace}
\def\KS    {\ensuremath{K^0_{\scriptscriptstyle S}}\xspace}

\def\Journal#1#2#3#4{{\it #1} {\bf #2}, #3 (#4)}

\long\def\inst#1{\par\nobreak\kern 4pt\nobreak
    {\it #1}\par\vskip 10pt plus 3pt minus 3pt}

\begin{document}
{\pagestyle{empty}

\begin{flushright}
SLAC-PUB-\SLACPubNumber \\
\babar-PROC-\BABARPubYear/\BABARProcNumber \\
October, 2006 \\
\end{flushright}

\par\vskip 4cm

\begin{center}
\Large \bf Study of Charm Baryons with the BaBar Experiment
\end{center}
\bigskip

\begin{center}
\large 
B.~Aa. Petersen\\
Stanford University\\
 Stanford, California 94305, USA\\
(for the \babar\ Collaboration)
\end{center}
\bigskip \bigskip

\begin{center}
\large \bf Abstract
\end{center}
We report on several studies of charm baryon production and decays by
the \babar\ collaboration. We confirm previous observations of the
$\Xi_c^{'0/+}$, $\Xi_c(2980)^+$ and $\Xi_c(3077)^+$ baryons, measure
branching ratios for Cabibbo-suppressed $\Lambda_c^+$ decays and use
baryon decays to study the properties of the light-quark baryons,
$\Omega^-$ and $\Xi(1690)^0$.

\vfill
\begin{center}
Contributed to the Proceedings of the 33$^{rd}$ International 
Conference on High Energy Physics, \\
7/26/2006---8/2/2006, Moscow, Russia
\end{center}

\vspace{1.0cm}
\begin{center}
{\em Stanford Linear Accelerator Center, Stanford University, 
Stanford, CA 94309} \\ \vspace{0.1cm}\hrule\vspace{0.1cm}
Work supported in part by Department of Energy contract DE-AC02-76SF00515.
\end{center}

\newpage

\section{Introduction}

The last few years have seen a revival of interest in charm
spectroscopy with more than a dozen new states being reported and
hundreds of new theoretical investigations being published.  The
\babar\ experiment\cite{BABAR} provides excellent opportunities to
observe and study new and old charm hadrons. It records $e^+e^-$ collisions at or
just below the \FourS\ resonance and with an integrated luminosity of
390\invfb, the recorded data sample contains more than one billion
charm hadron decays. About 10\% of these are charm baryons and we
report here on several charm baryon studies based on subsets of the
\babar\ data.

\section{Production of $\Xi_c^{'}$ Baryons}

The isospin doublet ($\Xi_c^{'0}$,$\Xi_c^{'+}$) was first observed by CLEO
in $e^+e^-$ continuum events\cite{bib:cleo:xic2575} and is
identified as the lightest $J^P=\frac{1}{2}^+$ $csq$ baryon doublet with 
a symmetric light-quark wave-function. This observation has now been
confirmed by \babar\ and its production measured in both $e^+e^-$ continuum
events and $B$ decays.\cite{bib:babar:xic2575}

\begin{figure}
\centerline{\psfig{file=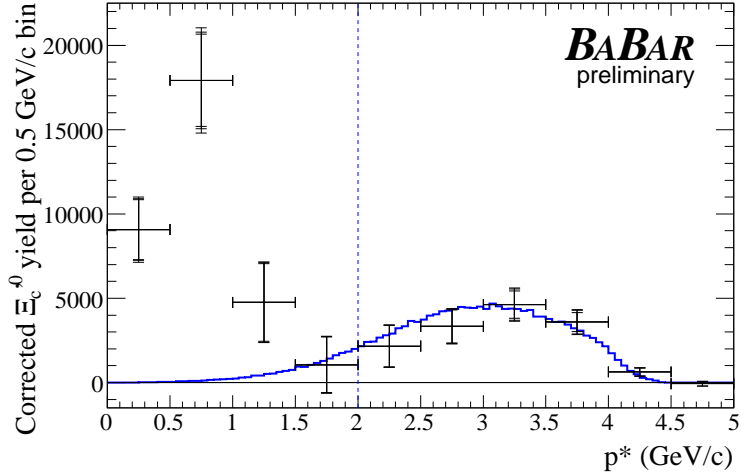,width=4in}}
\caption{ Efficiency-corrected, background-subtracted $p^*$ spectrum
  for $\Xi_c^{'0}$. The curve is the
  simulated continuum distribution; it is fitted to the data for
  $2.0 < p^* < 4.5\gevc$. }
\label{fig:xicprime}
\end{figure}

The \babar\ analysis uses 232\invfb of data. The $\Xi_c^{'0(+)}$
baryon is reconstructed in the only kinematically allowed decay mode:
$\Xi_c^{'0(+)}\to \Xi_c^{0(+)}\gamma$.  The $\Xi_c^{0(+)}$ is
reconstructed from the decay chain, $\Xi_c^{0(+)}\to\Xi^-\pip(\pip)$,
$\Xi^-\to\Lambda\pim$, $\Lambda\to p\pim$. Clear signals for
$\Xi_c^{'0(+)}$ are observed in the $\Xi_c^{0(+)}\gamma$ invariant
mass spectrum and yields are extracted from fits to the mass spectrum
in 0.5\gevc wide bins of $p^*$, the $\Xi_c^{'0(+)}$ momentum in
the $e^+e^-$ center-of-mass frame. The results for $\Xi_c^{'0}$ are
shown in Fig.~\ref{fig:xicprime} after correcting for reconstruction
efficiency. $\Xi_c^{'0(+)}$ baryons with $p^*>2\gevc$ come from
continuum production, while most baryons with $p^*<2\gevc$ are due to $B$
decays. Using a model of $\Xi_c^{'0(+)}$ continuum production, the yield for
$p^*>2\gevc$ is extrapolated to the full $p^*$ range. From this the
$B$ component is separated to obtain the branching fractions
$\mathcal{B}(B \rightarrow \Xi_c^{'+} X) \times \mathcal{B}(\Xi_c^+
\rightarrow \Xi^- \pi^+ \pi^+) = [ 1.69 \pm 0.17 \ \mathrm{(exp.)} \pm
0.10 \ \mathrm{(model)} ] \times 10^{-4}$ and $\mathcal{B}(B
\rightarrow \Xi_c^{'0} X) \times \mathcal{B}(\Xi_c^0 \rightarrow \Xi^-
\pi^+) = [ 0.67 \pm 0.07 \ \mathrm{(exp.)} \pm 0.03 \ \mathrm{(model)}
] \times 10^{-4}$.  This is the first observation of the $\Xi_c^{'}$
baryon in $B$ decays. The measured continuum production cross sections
are $ \sigma(e^+ e^- \rightarrow \Xi_c^{'+} X) \times
\mathcal{B}(\Xi_c^+ \rightarrow \Xi^- \pi^+ \pi^+) = 141 \pm 24 \
\mathrm{(exp.)} \pm 19 \ \mathrm{(model)} ~\mathrm{fb}$ and
$\sigma(e^+ e^- \rightarrow \Xi_c^{'0} X) \times \mathcal{B}(\Xi_c^0
\rightarrow \Xi^- \pi^+) = 70 \pm 11 \ \mathrm{(exp.)} \pm 6 \
\mathrm{(model)} ~\mathrm{fb}$. Comparing to a previous measurement of
$\Xi_c^{0}$ production,\cite{babar_xic} one observes that about one
third of the $\Xi_c^{0}$ produced in $B$ decays come from $\Xi_c^{'0}$
decays, while in continuum the fraction is only 18\%.

\section{$\Xi_c(2980)^+$ and $\Xi_c(3077)^+$}

Two new charm-strange baryons, $\Xi_c(2980)^+$ and $\Xi_c(3077)^+$,
were recently observed by BELLE in decays to
$\Lambda_c^+K^-\pip$.\cite{belle_xic2980} These are the first baryon
decays where the charm and strange quark are contained in separate
hadrons. \babar\ has confirmed this observation and studied the
resonant substructure in the decay.\cite{babar_xic2980}

For this study \babar\ uses 316\invfb of data. The $\Lambda_c^+$
baryons are reconstructed in the decay mode $\Lambda_c^+\to pK^-\pip$
and combined with a second kaon and pion candidate. The invariant mass
of the $\Lambda_c^+\pip$ pair are plotted versus the mass of the
$\Lambda_c^+K^-\pip$ candidate in Fig.~\ref{fig:xicstarscatter}. 
Horizontal bands from the $\Sigma_c(2520)^{++}$ and $\Sigma_c(2455)^{++}$ resonances
are observed and enhancements around 2970 and 3077\mevcc in $M(\Lambda_c^+K^-\pip)$ can also be seen.

\begin{figure}
\centerline{\psfig{file=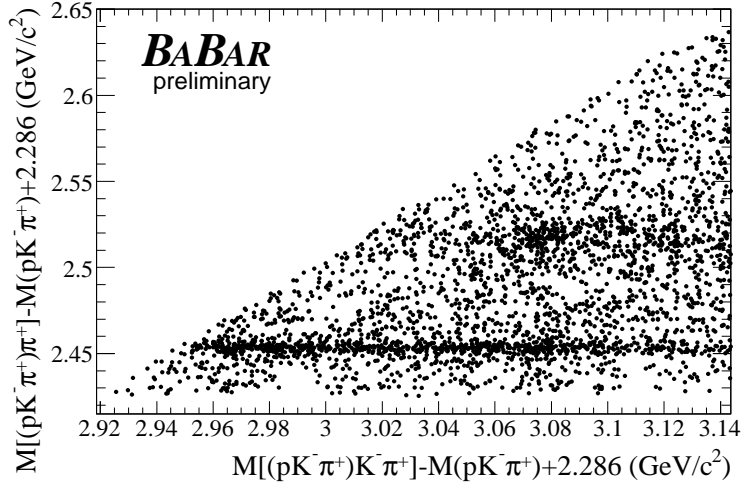,width=4in}}
\caption{Two-dimensional scatter plot of $M(\Lambda_c^+\pip)$
vs. $M(\Lambda_c^+K^-\pip)$ for $pK^-\pi^+K^-\pi^+$ candidates. }
\label{fig:xicstarscatter}
\end{figure}

\begin{figure}
\centerline{\psfig{file=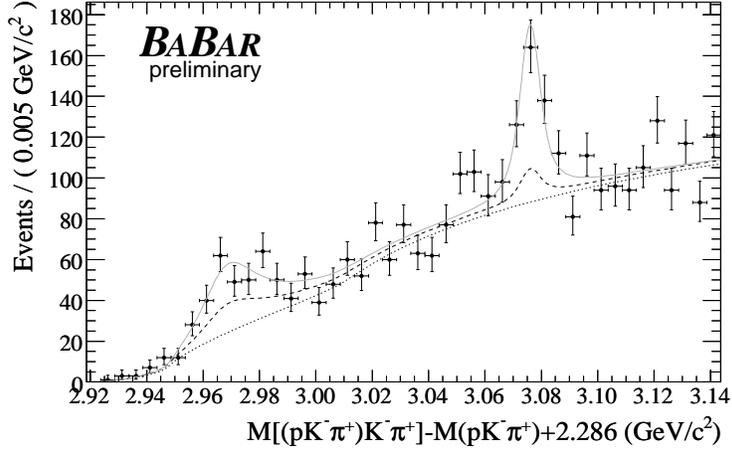,width=4in}}
\caption{Invariant mass distribution of $\Lambda_c^+K^-\pip$ for data
(points with error bars) and the fitted PDF (curves). The solid curve
shows the total fit PDF. The dotted curve shows the sum of the
background components, while the dashed curve on top shows the sum of
the non-resonant signal components.}
\label{fig:xicstarfit}
\end{figure}

To extract the yields of the $\Xi_c(2980)^+$ and $\Xi_c(3077)^+$
baryons and determine how much decays through a $\Sigma_c^{++}$
resonance, a two-dimensional unbinned likelihood fit is performed on
the events shown in Fig.~\ref{fig:xicstarscatter}. The fit contains
both resonant $\Sigma_c^{++}$ components and non-resonant components
for both the $\Xi_c(2980)^+$ and $\Xi_c(3077)^+$ signals and the
background. Phase-space suppression factors are including in the
signal PDFs to account for the nearby kinematic thresholds.
\begin{table}
\caption{Yields for the separate resonant and non-resonant (NR) decays.\label{tab:XicstarYields}}
\begin{center}
{\begin{tabular}{@{}lc@{}}
\hline
    & Yield (Events)    \\ \hline
$\Xi_c(2980)^+\rightarrow\Sigma_c(2455)^{++}K^-$     & $132\pm31\pm5$      \\
$\Xi_c(2980)^+\rightarrow\Lambda_c^+K^-\pi^+$ (NR)   & $152\pm37\pm45$     \\ \hline
$\Xi_c(3077)^+\rightarrow\Sigma_c(2455)^{++}K^-$     & $87\pm20\pm4$       \\
$\Xi_c(3077)^+\rightarrow\Sigma_c(2520)^{++}K^-$     & $82\pm23\pm6$       \\
$\Xi_c(3077)^+\rightarrow\Lambda_c^+K^-\pi^+$ (NR)   & $35\pm24\pm16$      \\ 
\hline
\end{tabular}}
\end{center}
\end{table}

\begin{table}
\caption{Comparison of masses and widths for $\Xi_c(2980)^+$ and $\Xi_c(3077)^+$, measured by \babar\ and BELLE in the $\Lambda_c^+K^-\pi^+$ final state.\label{tab:Xicstarmasswidth}}
\begin{center}
{\begin{tabular}{@{}lcc@{}}
\hline
 & Mass (\mevcc)   & Width (\mev)    \\ \hline
\babar\  & $2967.1\pm1.9\pm1.0$    & $23.6\pm2.8\pm1.3$  \\
BELLE    & $2978.5\pm2.1\pm2.0$    & $43.5\pm7.5\pm7.0$  \\ \hline
\babar\  & $3076.4\pm0.7\pm0.3$    & $6.2\pm1.6\pm0.5$   \\
BELLE   & $3076.7\pm0.9\pm0.5$    & $6.2\pm1.2\pm0.8$   \\
\hline
\end{tabular}}
\end{center}
\end{table}

The fitted PDF is overlaid on the data in Fig.~\ref{fig:xicstarfit}
and the fit yields are given in Table~\ref{tab:XicstarYields}. The
statistical significance of both the $\Xi_c(2980)^+$ and the
$\Xi_c(3077)^+$ signals exceed $7\,\sigma$. The $\Sigma_c^{++}$
resonances are seen to be dominant in the $\Xi_c(3077)^+$ decay, while
only about 50\% of the $\Xi_c(2980)^+$ decays are through
$\Sigma_c(2455)^{++}$.  The mass and width of the two resonances are
also obtained in the fit and are compared to the results from BELLE in
Table~\ref{tab:Xicstarmasswidth}.  The parameters for the
$\Xi_c(3077)^+$ baryon agree, but for the $\Xi_c(2980)^+$ the \babar\ mass
and width are lower. This might be due to different
treatment of the phase-space suppression.

\section{Cabibbo-Suppressed $\Lambda_c^+$ Decays}

Only a few Cabibbo-suppressed $\Lambda_c^+$ decays have been measured
and most have poor precision. \babar\ has performed
a precise measurement of two Cabibbo-suppressed decay modes and
searched for two new decays.\cite{babar_lc} 

The analysis uses 125\invfb of data.  It combines $\Lambda$ and
$\Sigma^0$ candidates with $K^+$ candidates to reconstruct
$\Lambda_c^+\ra \Lambda K^+$ and $\lcpl \ra \Sigma^{0} K^+$
decays. Large signals are seen in both decay modes and used together
with signals from the Cabibbo-favored $\Lambda_c^+\ra \Lambda \pip$ and
$\lcpl \ra \Sigma^{0} \pip$ decays to obtain the branching
ratios:
\begin{eqnarray*}
\frac{\brc(\Lambda_c^+ \ra \Lambda K^+)}{\brc(\lcpl \ra \Lambda \pip )} &= &0.044~\pm~0.004~\pm~0.003,\\
\frac{\brc(\lcpl \ra \Sigma^{0} K^+)}{\brc(\lcpl \ra \Sigma^{0} \pip )} &= &0.039~\pm~0.005~\pm~0.003,
\end{eqnarray*}
where the first uncertainty is statistical and the second
systematic. These are a significant improvement over previous
measurements\cite{belle_lc} and in agreement with quark-model
predictions.\cite{Khana}

Searches for the four-body decays $\lcpl \ra \Lambda
K^+ \pip \pim$ and $\lcpl \ra \Sigma^{0} K^+ \pip \pim$ are also performed. After
removing decays with intermediate resonances ($\Lambda K^+\KS$ and
$\Xi^-K^+\pip$), no significant signals are observed and the following
upper limits at 90\% confidence level are obtained:
\begin{eqnarray*}
\frac{\brc(\lcpl \ra \Lambda K^+ \pip \pim)}{\brc(\lcpl \ra \Lambda \pip )} &<& 4.1 \times ~10^{-2},\\
\frac{\brc(\lcpl \ra \Sigma^{0} K^+ \pip \pim)}{\brc(\lcpl \ra \Sigma^{0} \pip )} &<& 2.0 \times ~10^{-2}.
\end{eqnarray*}
The measurements are the first limits on these decay modes.

\section{Measurement of the $\Omega^-$ Spin}

The quark model successfully predicted\cite{gellMann} the existence of
the $\Omega^-$ baryon. It also predicts its spin to be $3/2$, but
until now experiments have only established that its spin is larger
than $1/2$. \babar\ has studied its spin by using
$\Xi_c^0\to\Omega^-\pip, \Omega^-\to\Lambda K^-$ decays reconstructed
in 116\invfb of data.\cite{babar_omega} The distribution of the helicity angle
$\theta_{h}$, defined as the angle between the $\Lambda$ and the
$\Xi_c^0$ in the $\Omega^-$ rest-frame, depends on the spin $J$ of the $\Omega^-$ and the
$\Xi_c^0$. Assuming $J_{\Xi_c^0}=1/2$, the distributions for $J_\Omega=1/2$, $3/2$ and
$5/2$, respectively are:
\begin{eqnarray}
{dN}/{d\cos \theta_{ h}}&\propto& 1\\
{dN}/{d\cos \theta_{ h}}&\propto& 1 + 3\,{\cos}^2\theta_{h}\\
{dN}/{d\cos \theta_{ h}}&\propto&  1-2\,{\cos}^2\theta_{h}+5\,{\cos}^4\theta_{h}
\end{eqnarray}
Parity violation in the $\Xi_c^0$ and $\Omega^-$ decays could
give an additional term with odd powers of $\cos\theta_{h}$, but no evidence
for that is seen in data.

\begin{figure}
\centerline{\psfig{file=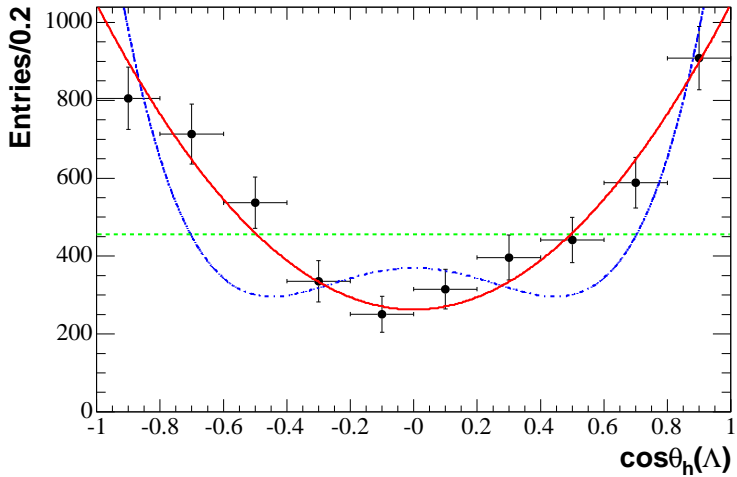,width=4in}}
\caption{The efficiency-corrected cos$\theta_h(\Lambda)$ distribution
for $\Xi_c^0 \rightarrow \Omega^- K^+$ data.  The curves 
represents the expected distribution for $J_{\Omega}=1/2$ (dashed), $J_{\Omega}=3/2$ (solid) 
and $J_{\Omega}=5/2$ (dashed-dotted).}
\label{fig:omegafit}
\end{figure}

Figure~\ref{fig:omegafit} shows the helicity angle distribution with the
three spin hypothesis overlaid. The $J_{\Omega}=3/2$ hypothesis is in clear
agreement with data, while the confidence levels for the
$J_{\Omega}=1/2$ and $J_{\Omega}=5/2$ hypotheses are $1\times 10^{-17}$ and
$3\times10^{-7}$, respectively. This confirms that the $\Omega^-$ baryon has the
expected spin.

\section{Study of the $\Xi(1690)^0$ Baryon}

The existence of the $\Xi(1690)^0$ baryon has been known for many
years, but relatively little information is available on it. BELLE
has observed it as an intermediate resonance in
$\Lambda_c\to(\Lambda\KS)K^+$ decays.\cite{belle_lc} \babar\ uses
this decay mode in a larger data sample (200\invfb) to measure the
mass, width and spin of the $\Xi(1690)^0$ baryon.\cite{babar_xi1690}

\babar\ reconstructs $2750\pm 300$ $\Lambda_c\to\Lambda\KS K^+$ decays
and the $\Xi(1690)^0$ is clearly visible in invariant mass of the
$\Lambda\KS$ pair (see Fig.~\ref{fig:xi1690fit}). The remaining
$\Lambda_c^+$ decays cannot be described as a non-resonant
contribution. Instead they appear to be decays through
$\Lambda a_0(980)^+$, where $a_0(980)^+\to\KS K^+$. To measure the mass
and width of the $\Xi(1690)^0$, the mass projections of the
$\Lambda\KS K^+$ Dalitz plot are fitted to a coherent sum of
$\Lambda a_0(980)^+$ and $\Xi(1690)^0K^+$ decays. The fit result is
overlaid in Fig.~\ref{fig:xi1690fit} and gives
\begin{eqnarray*}
m(\Xi(1690)) &=& 1684.7{\pm 1.3}\,^{+2.2}_{-1.6} \mevcc,\\
\Gamma(\Xi(1690))&=&8.1_{-3.5}^{+3.9}\,^{+1.0}_{-0.9}\mev,
\end{eqnarray*}
where the first uncertainty is statistical and the second systematic,
primarily related to the interference contribution. Both the mass and
the width measurements are significant improvements over previous
results.

\begin{figure}
\centerline{\psfig{file=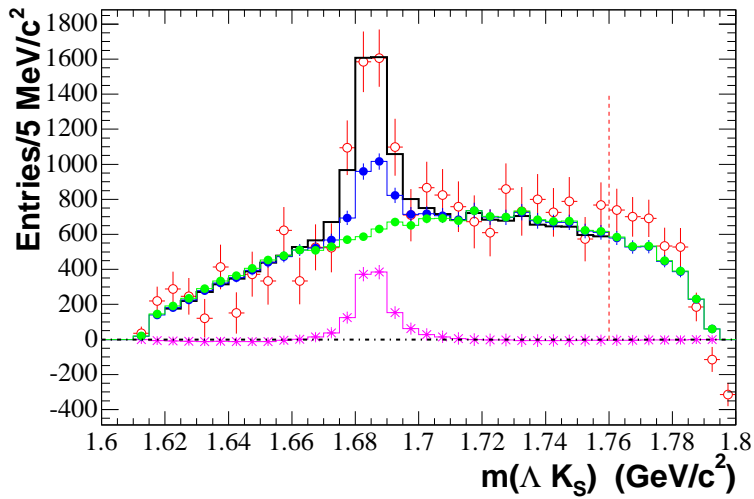,width=4in}}
\caption{The $\Lambda K_S$ invariant mass projection in data (open
circles) with the fit overlaid (black line). Superimposed is the
distribution for the $\Lambda a_0(980)^+$ contribution (light-colored
dots), the interference term (stars) and their sum (dark-colored
dots).}
\label{fig:xi1690fit}
\end{figure}

The spin of the $\Xi(1690)^0$ baryon can be studied analogously to the
$\Omega^-$ spin.  Assuming the $\Lambda_c^+$ baryon has spin $1/2$, the
$\Xi(1690)^0$ is found to be fully consistent with spin $1/2$, while
the spin $3/2$ and $5/2$ hypotheses only have confidence levels of
0.02 and 0.01, respectively.  This is the first spin measurement for the
$\Xi(1690)^0$ baryon.

\section{Summary}

The large sample of charm hadrons recorded by the \babar\ experiment
has been used to study the properties of several charm
baryons. Furthermore, the decays of charm baryons are used as a clean
source of light-quark baryons whose spin, mass and width are
measured.

\end{document}